\documentclass[12pt]{article}
\pdfoutput=1
\usepackage[latin1]{inputenc}
\usepackage{amsmath}
\usepackage{amsfonts}
\usepackage{amssymb}
\usepackage{graphicx}
\usepackage{geometry}
\usepackage{amssymb,epsfig,subfigure}
\usepackage{hyperref}

\def\simge{
    \mathrel{\rlap{\raise 0.511ex 
        \hbox{$>$}}{\lower 0.511ex \hbox{$\sim$}}}}

\def\simle{
    \mathrel{\rlap{\raise 0.511ex 
        \hbox{$<$}}{\lower 0.511ex \hbox{$\sim$}}}}

\makeatletter
\renewcommand\section{\@startsection {section}{1}{\z@}%
                                 {-3.5ex \@plus -1ex \@minus -.2ex}
                                   {2.3ex \@plus.2ex}%
                                   {\normalfont\large\bfseries}}
\renewcommand\subsection{\@startsection{subsection}{2}{\z@}%
                                   {-3.25ex\@plus -1ex \@minus -.2ex}%
                                     {1.5ex \@plus .2ex}%
                                     {\normalfont\bfseries}}
\renewcommand\subsubsection{\@startsection{subsubsection}{3}{\z@}%
                                   {-3.25ex\@plus -1ex \@minus -.2ex}%
                                     {1.5ex \@plus .2ex}%
                                     {\normalfont\itshape}}
\makeatother

\def\pplogo{\vbox{\kern-\headheight\kern -29pt
\halign{##&##\hfil\cr&{\ppnumber}\cr\rule{0pt}{2.5ex}&\ppdate\cr}}}
\makeatletter
\def\ps@firstpage{\ps@empty \def\@oddhead{\hss\pplogo}%
  \let\@evenhead\@oddhead 
}
\def\maketitle{\par
 \begingroup
 \def\thefootnote{\fnsymbol{footnote}}
 \def\@makefnmark{\hbox{$^{\@thefnmark}$\hss}}
 \if@twocolumn
 \twocolumn[\@maketitle]
 \else \newpage
 \global\@topnum\z@ \@maketitle \fi\thispagestyle{firstpage}\@thanks
 \endgroup
 \setcounter{footnote}{0}
 \let\maketitle\relax
 \let\@maketitle\relax
 \gdef\@thanks{}\gdef\@author{}\gdef\@title{}\let\thanks\relax}
\makeatother

\numberwithin{equation}{section}

\newcommand{\be}{\begin{equation}}
\newcommand{\bea}{\begin{eqnarray}}
\newcommand{\ee}{\end{equation}}
\newcommand{\eea}{\end{eqnarray}}
\newcommand\beq{\begin{equation}}
\newcommand\eeq{\end{equation}}

\def\be{\begin{equation}}
\def\ee{\end{equation}}
\def\ba#1\ea{\begin{align}#1\end{align}}
\def\bg#1\eg{\begin{gather}#1\end{gather}}
\def\bm#1\em{\begin{multline}#1\end{multline}}
\def\bmd#1\emd{\begin{multlined}#1\end{multlined}}

\begin{document}

\setcounter{page}0
\def\ppnumber{\vbox{\baselineskip14pt
}}
\def\ppdate{\footnotesize{SU-ITP-14/09}} \date{}

\author{Peter W.\ Graham$^a$, Bart Horn$^b$, Surjeet Rajendran$^a$, Gonzalo Torroba$^a$\\
[7mm]
{\normalsize \it $^a$ Stanford Institute for Theoretical Physics, Department of Physics, }\\
{\normalsize  \it  Stanford University, Stanford, CA 94305, USA }\\
[3mm]
{\normalsize \it $^b$Theory Group and Institute for Strings, Cosmology, and Astroparticle Physics,}\\
{\normalsize  \it Columbia University, New York, NY, 10027, USA}\\
[3mm]}

\bigskip
\title{\bf  Exploring eternal stability with the simple harmonic universe
\vskip 0.5cm}
\maketitle

\begin{abstract}
We construct nonsingular cyclic cosmologies that respect the null energy condition, have a large hierarchy between the minimum and maximum size of the universe, and are stable under linearized fluctuations. The models are supported by a combination of positive curvature, a negative cosmological constant, cosmic strings and matter that at the homogeneous level behaves as a perfect fluid with equation of state $-1<w<-1/3$. We investigate analytically the stability of the perturbation equations and discuss the role of parametric resonances and nonlinear corrections. Finally, we argue that Casimir energy contributions associated to the compact spatial slices can become important at short scales and lift nonperturbative decays towards vanishing size. This class of models (particularly in the static limit) can then provide a useful framework for studying the question of the ultimate (meta)stability of an eternal universe.
\end{abstract}
\bigskip
\newpage

\tableofcontents

\vskip 1cm

\section{Introduction}\label{sec:intro}

The construction of nonsingular cosmological solutions is a central problem in General Relativity (GR), both for its implications on the mathematical structure of the theory and for potential applications to describing our universe. This subject combines deep physical questions --did our Universe have a beginning, or is it eternal?-- with precise theorems regarding general properties of solutions. The singularity theorems in GR\footnote{See e.g.~\cite{Hawking:1973uf} for a detailed discussion.} imply that, under fairly general assumptions, cosmological solutions must be singular. Even inflating solutions have been shown to be geodesically incomplete in the past~\cite{Borde:2001nh}. In this context, it is important to try to extend the validity of these theorems, or to find nonsingular models and understand how the theorems are evaded. Moreover, the existence of such models would have interesting consequences for inflation, by providing possible past-completions or by decaying into realistic cosmologies.

A key assumption in the original singularity theorems is the strong energy condition (SEC) $T_{\mu\nu} v^\mu v^\nu \ge \frac{1}{2}T v_\mu v^\mu$, where $T_{\mu\nu}$ is the energy-momentum tensor, and $v_{\mu}$ is a timelike vector.  The SEC is, however, known to be violated in many consistent systems, and therefore one way to proceed is to relax this condition. It is reasonable to impose instead the (weaker)
null energy condition (NEC) $T_{\mu\nu} v^\mu v^\nu \ge \frac{1}{2}T v_\mu v^\mu$, where this time $v_\mu$ is a null vector. This condition appears to be satisfied macroscopically in our universe. Furthermore, for the purpose of performing a self-consistent GR study, it is important to demand that
the solutions be weakly curved, with all scales much larger than the Planck length. Under these conditions, positive curvature is the main ingredient that allows the circumvention the singularity theorems.\footnote{Another interesting possibility, recently studied in~\cite{Balasubramanian:2014jaa}, is to use translation-breaking effects.}

Ref.~\cite{usfirst} (see also~\cite{Harrison:1967zz, Kardashev:1990, Dabrowski:1995ae} for earlier related work) constructed a family of nonsingular solutions that satisfy the NEC and are everywhere weakly curved. These are bouncing universes that arise from the competition between positive curvature, a negative cosmological constant, and a fluid with equation of state $-1<w<-1/3$. In particular, for $w=-2/3$ the FRW equations can be solved analytically, leading to a scale factor that oscillates harmonically ---the ``simple harmonic universe'' (SHU). Our goal in this work is to continue the analysis of this solution, and improve it in different directions. 

First, the model of~\cite{usfirst} is linearly stable only if the ratio $a_-/a_+$ between the smallest and largest sizes of the universe is not too small. In \S \ref{sec:multifluid} we will show that by adding extra matter sources (cosmic strings) it is possible to
find stable solutions with a large hierarchy $a_-/a_+\ll 1$. This is an important step in order to construct realistic bouncing cosmologies. 

Secondly, we will perform in \S \ref{sec:stability} a detailed study of the linearized fluctuations, investigating analytically their stability and discussing the role of parametric resonances. This will establish the linearized stability of our solutions over a large range of parameter space. This analysis extends to the quantum level by fixing the size of the initial perturbation by the uncertainty principle.
Nonlinearities could cut off this behavior, leading to a finite but plausibly exponentially long lifetime. It is, however, possible that the nonlinearities do not lead to instabilities -- the nearly static limit is argued to be very different from usual thermodynamic systems, and it is as of yet an open question whether this class of systems may be eternal even after the inclusion of nonlinear effects.

A universe that is stable at the linear and nonlinear level may still be subject to nonperturbative decays, making it metastable (but exponentially long-lived). \S \ref{sec:nonpert} focuses on the possible nonperturbatively suppressed decays of our cyclic cosmologies. The simplest example of an instanton where the universe tunnels to nothing was found in~\cite{Mithani:2011en}. We argue that this instability can be eliminated by introducing additional sources of energy density such as  Casimir energy corrections associated with compact spatial slices. 
While we do not provide a definite answer to whether these cosmologies could then be truly eternal, and there are potentially other nonperturbative decay channels that need to be addressed, we believe that this class of models provides a controlled setup where the issue of the beginning of time could be further studied. Our conclusions and future directions are presented in \S \ref{sec:concl}.

\section{Cyclic universes with large bounces}\label{sec:multifluid}

It is interesting to ask if a cyclic cosmology can accommodate a large ratio between the minimum and maximum sizes of the universe which could reproduce, for instance, the hierarchy between the GUT (or Planck) scale, and the size of our Hubble patch. However, it was shown in~\cite{usfirst} that in models with only one type of fluid, the linearized fluctuations with momenta $l \lesssim \sqrt{a_+/a_-}$ grow exponentially. Intuitively, the Hubble antifriction produced by the evolution near the minimum dominates on average over the friction contribution from the maximum, effectively inducing a tachyonic mass. For bounces with  $a_+/a_- \gg 1$, this leads to unacceptable instabilities in a large number of modes. 

We will now show that it is possible to construct cyclic models that do not suffer from such instabilities by adding one more matter source -- {e.g.}\ a frustrated network of cosmic strings. The additional source of energy density enables the frequency of oscillation of the universe to be parametrically decoupled from the mass of the linearized fluctuations, which is set by the curvature of the space-time. This freedom was absent in \cite{usfirst} and it permits  the frequency of the oscillations to be small compared to the mass of the fluctuations, enabling stability.

\subsection{General equations and energetic considerations}\label{subsec:general}

Let us begin our analysis by determining how a cyclic solution can arise from the competition of different matter and energy sources. 

We consider an FRW metric with positive curvature,
\be\label{eq:metric1}
ds^2 = -dt^2 +a(t)^2 \left(\frac{dr^2}{1- K r^2}+ r^2(d\theta^2+ \sin^2 \theta\,d\phi^2) \right)\,.
\ee
Here the curvature $K=1$ and the spatial slices are 3-spheres; we will find it useful to keep the dependence on $K$ explicit, to account more simply below for the effects of cosmic strings. The FRW equations in the presence of a cosmological constant $\Lambda$ and additional matter sources are
\bea\label{eq:FRW}
\frac{\ddot a}{a} &=& - \frac{4\pi}{3} G_N (\rho + 3 p)+ \frac{\Lambda}{3} \nonumber\\
\left( \frac{\dot a}{a}\right)^2&=& \frac{8\pi}{3} G_N \,\rho - \frac{K}{a^2}+ \frac{\Lambda}{3}\,.
\eea
For simplicity, we will restrict to matter sources that behave macroscopically (but not microscopically, as we discuss in \S \ref{subsec:stability}) as perfect fluids, 
\be
p = w \rho\;,\;\rho= \frac{c_w}{a^{3+3w}}\,.
\ee

To proceed, it will be useful to recast the second FRW equation in (\ref{eq:FRW}) as the vanishing of the GR Hamiltonian\footnote{Recall that the constraint $H=0$ is imposed by the $g_{00}$ Lagrange multiplier, which we have here gauge-fixed to $g_{00}=-1$.}
\be\label{eq:GRH}
H =- \frac{G_N}{3\pi a}\left(\pi_a^2 + V(a) \right)
\ee
where $\pi_a$ is the canonical momentum
\be
\pi_a = \frac{\partial L}{\partial \dot a} = -\frac{3 \pi}{2 G_N} a \dot a
\ee
and the effective potential
\be\label{eq:Va}
V(a) = \left(\frac{3\pi}{2G_N} \right)^2\,a^2 \left(K - \frac{8\pi}{3}G_N c_w \,a^{-3w-1}- \frac{1}{3}\Lambda a^2 \right)\,.
\ee
Given this three-term structure, it is clear what sources are needed for a periodic solution. At the minimum and maximum, $\pi_a=0$, which means $V(a)$ should admit two roots $a_\pm$; furthermore $V(a)<0$ between these roots so that $a$ oscillates between $a_-$ and $a_+$ as determined by (\ref{eq:GRH}). We therefore need
\be\label{eq:general-matter}
-1 < w < - \frac{1}{3}\;,\;\Lambda<0\,.
\ee
The evolution near the minimum is dominated by curvature (which tries to make the universe expand) and the fluid (which pushes towards contraction), while near the maximum the negative cosmological constant balances against the fluid. Denoting the period of oscillation by $\tau$ we also obtain, from the integral of the second FRW equation,
\be
\frac{\tau}{2} = \int_{a_-}^{a_+} \frac{da}{\left(\frac{8\pi}{3} G_N \,c_w a^{3|w|-1}  + \frac{\Lambda}{3} a^2-K \right)^{1/2}}\,.
\ee

Before discussing models with a large hierarchy $a_- \ll a_+$, we point out that the case $a_- \sim a_+$ is also of conceptual interest. We can tune the parameters $c_w$ and $\Lambda$ in (\ref{eq:Va}) so that the two roots nearly coincide, $a_- \approx a_+$, and this corresponds to the static limit of our cyclic cosmologies. Unlike the Einstein static universe, this class of solutions does not suffer from homogeneous instabilities, partly because it does not require tuning the initial conditions. Therefore, the models with positive curvature and matter content (\ref{eq:general-matter}) provide a framework where many of the questions studied in the Einstein static universe can be addressed without the danger of linearized instabilities.\footnote{For work on the Einstein static universe, its stability and applications to inflationary cosmology see e.g.~\cite{Einstein}.}  In particular, we will return to the question of the ultimate (meta)stability of such static cosmologies in \S \ref{sec:nonpert}.

Let us now find the conditions for a large hierarchy. Assuming $a_- \ll a_+$,  the minimum is approximately determined by balancing in (\ref{eq:Va}) the curvature and fluid contributions,
\be\label{eq:amin}
a_- \approx \left(\frac{3}{8\pi} \frac{K}{G_N c_w} \right)^{\frac{1}{3|w|-1}}\,,
\ee
while for the maximum we obtain
\be\label{eq:amax}
a_+ \approx \left(8 \pi \frac{G_Nc_w}{|\Lambda|} \right)^{\frac{1}{3-3|w|}}\,.
\ee
Therefore, choosing the cosmological constant and density such that
\be\label{eq:cond}
|\Lambda|^{3|w|-1} \ll \frac{(8\pi G_N c_w)^2}{K^{3-3|w|}}\,
\ee
obtains $a_- \ll a_+$. Furthermore, as long as the matter density
\be
c_w \ll \frac{K}{(8 \pi G_N)^\frac{1+3|w|}{2}}\,,
\ee
the smallest size of the universe will be much larger than the Planck length, guaranteeing that classical GR remains a good approximation. 

We also note that these results generalize easily to multifluid models, replacing $\rho = \sum_i  c_i \,a^{-3(1+w_i)}$ in the expressions before. In particular, let us see what happens when we combine a fluid with equation of state $-1 < w < - 1/3$ with a density of cosmic strings, $w=-1/3$. Since the contribution of strings is $\rho=c_{str}/a^2$, their
effect is simply encoded in a redefinition of the curvature,
\be\label{eq:Keff}
 K_{eff} = K - \frac{8\pi}{3} G_N c_{str}\,,
\ee
and the previous equations still apply after replacing $K \to K_{eff}$ (we can now set the background curvature $K=1$ inside $K_{eff}$). Increasing the density of strings decreases the effective curvature, and by tunning $G_N c_{str} \sim 3/8\pi$ it is possible to nearly cancel $K_{eff}$. Now, from (\ref{eq:amin})--(\ref{eq:cond}) we see that taking $K_{eff} \ll 1$ while keeping the other parameters fixed leads to a hierarchy $a_-/a_+ \ll 1$. Moreover, we will show below in \S \ref{subsec:stability} that a smaller $K_{eff}$ gives a larger mass from the angular momentum on the sphere. Therefore, the addition of cosmic strings can simultaneously lead to solutions with large hierarchies between the minimum and maximum size, and to a stabilization of low-lying modes that could otherwise be unstable.

\subsection{An analytic solution with domain walls and cosmic strings}\label{subsec:DWCS}

We now illustrate our previous general results by studying the special (and solvable) case of domain walls (or any $w = -2/3$ fluid) plus cosmic strings.

The solution to the FRW equations is a simple harmonic universe, albeit now with a nontrivial $K_{eff}$,
\be\label{eq:aSHU}
a(t) = \frac{1}{\omega} \frac{\sqrt{K_{eff}}}{\sqrt{\gamma}} \left(1+ \sqrt{1-\gamma} \,\cos\,\omega t \right)\,.
\ee
The frequency of oscillations is
\be
\omega = \sqrt{\frac{|\Lambda|}{3}}\;,
\ee
and we have introduced the combination
\be
\gamma =  \frac{3 K_{eff}|\Lambda|}{(4\pi G_N c_{dw})^2}\,.
\ee

When $\gamma \sim 1$ the oscillatory term in (\ref{eq:aSHU}) vanishes, and we obtain a static solution with $a_-=a_+=\frac{1}{\omega} \sqrt{K_{eff}}$. On the other hand, the limit $\gamma \ll 1$ gives
\be
a_+ \approx \frac{2}{\omega}\frac{\sqrt{K_{eff}}}{\sqrt{\gamma}}\;,\;a_-\approx \frac{1}{2\omega} \sqrt{\gamma K_{eff}}\,,
\ee
in agreement with (\ref{eq:amin}) and (\ref{eq:amax}) for $w=-2/3$ and $K \to K_{eff}$. This leads to a hierarchy
\be
\frac{a_-}{a_+} \approx  \frac{\gamma}{4} \ll 1\,.
\ee

In the rest of this work we will often focus on this special case, which allows for analytic results. However, we would like to stress that our results will apply to the more general class of models with fluid $-1<w<-1/3$. For short, we will refer to the general family of solutions as the ``simple harmonic universe'', in a slight abuse of notation.

\subsection{Stability of linearized perturbations}\label{subsec:stability}

As we mentioned before, in the limit $a_- \ll a_+$, cyclic universes supported by curvature, a negative cosmological constant and a matter source with $-1<w<-1/3$ are unstable under linear perturbations. We will now argue that adding an appropriate density of cosmic strings solves this problem.

The key point is that in the presence of cosmic strings the effective curvature becomes an adjustable parameter [see (\ref{eq:Keff})], and using this extra degree of freedom we can keep the ratio $a_-/a_+$ and the period of oscillation $\tau$ fixed, while the overall scale $\sqrt{a_- a_+}$ is decreased.\footnote{For instance, in the analytic solution of \S \ref{subsec:DWCS}, this is achieved by fixing $\Lambda$ and $\gamma$, and decreasing $K_{eff}$ and $c_{dw}$ in a correlated way. For more general $w$, the cosmological constant also needs to be changed together with the other parameters.} Since the masses of linear fluctuations are inversely proportional to this scale, a sufficiently small $K_{eff}$ can overcome the tachyonic contribution from Hubble antifriction, leading to linearly stable perturbations. 

Let us illustrate this effect of cosmic strings in the analytic SHU model of \S \ref{subsec:DWCS}. It is convenient to change to conformal time,
$d\eta^2 = dt^2/a(t)^2$, which yields a scale factor
\be
a(\eta) = \frac{\sqrt{\gamma K_{eff}}}{\omega} \frac{1}{1-\sqrt{1-\gamma}\,\cos (\sqrt{K_{eff}}\eta)}\,.
\ee
As a simple and generic example of linear perturbation, we analyze a probe scalar field $\phi$ in the FRW background,
\be\label{eq:massless}
\frac{d^2 \phi}{d\eta^2} + 2 \frac{da/d\eta}{a} \,\frac{d\phi}{d\eta} - \nabla^2_{S^3}\phi=0\,.
\ee
Redefining
\be
\hat \eta \equiv \sqrt{K_{eff}}\eta
\ee
obtains
\be\label{scalareq}
\frac{d^2 \phi}{d \hat \eta^2} + 2\frac{da/d \hat\eta}{a} \, \frac{d\phi}{d\hat\eta} - \frac{1}{K_{eff}}\,\nabla^2_{S^3}\phi=0\,.
\ee
In this form, the only dependence on $K_{eff}$ appears in the $S^3$ Laplacian term. It is now clear that $K_{eff}\ll 1$ enhances the positive contribution to the mass squared for modes with nonzero momentum.

\begin{figure}[h!]
\begin{center}
\includegraphics[width=0.4\textwidth]{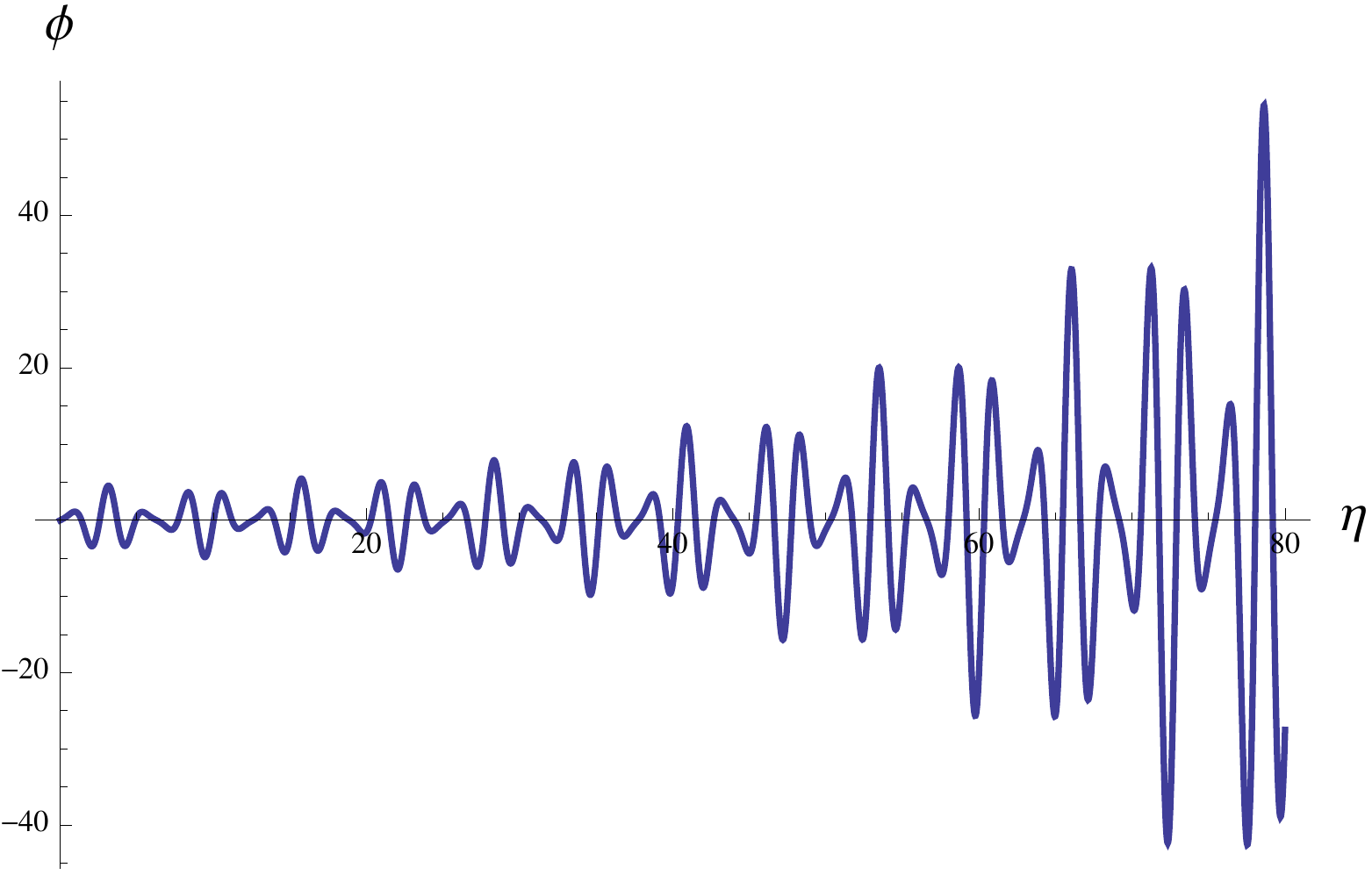}
\includegraphics[width=0.4\textwidth]{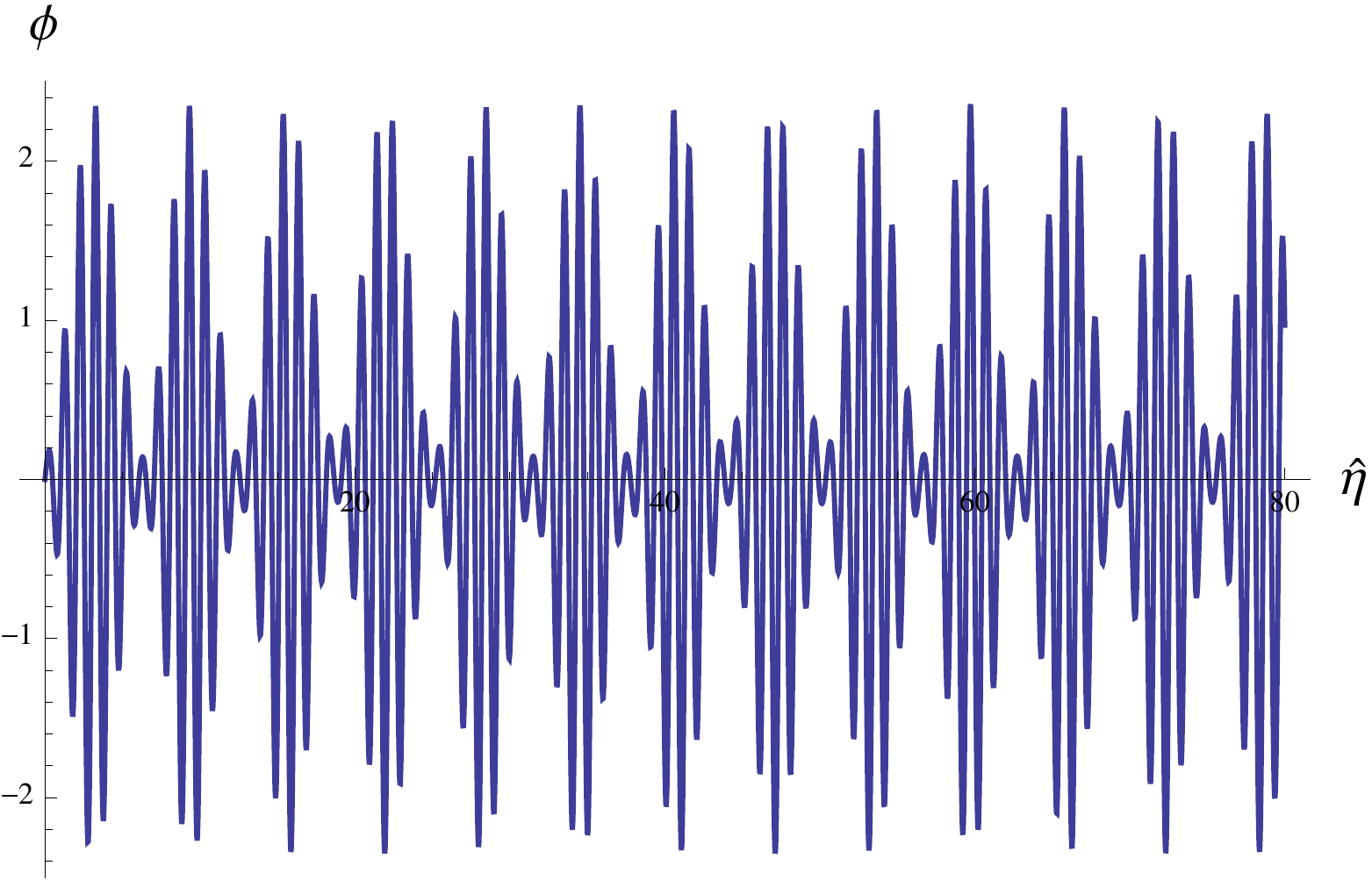}
\end{center}
\caption{\small{Evolution of the $l=2$ harmonic mode in a SHU with $\gamma=0.225$. In the first plot there are no cosmic strings, and this mode grows exponentially. The second plot shows how the mode becomes stable in the presence of a density of cosmic strings with $K_{eff}=0.2$.}}\label{fig:stable}
\end{figure}

We can simplify this equation further by expanding in spherical harmonics and redefining the field to cancel the term linear in derivatives. Doing this for the $l^{th}$ harmonic obtains
\be\label{eq:psieom}
\psi_l''+ m_l^2 \psi_l = 0\;,\;m_l^2 = \frac{l(l+2)}{K_{eff}}- \frac{a''}{a}\,.
\ee
The largest tachyonic contribution from the time evolution of the scale factor occurs near $\hat \eta \approx \sqrt{\frac{5\gamma}{3}}$ and is of order $\frac{a''}{a}\approx \frac{9}{8\gamma}$. Thus, tuning the density of cosmic strings $c_{str}$ in (\ref{eq:Keff}) such that
\be\label{stabilitylargebounce}
K_{eff} \lesssim \gamma
\ee
the linearized perturbations for generic values of the momentum $l$ will be stable even in the regime $\gamma \ll 1$. We show an example of this in Figure \ref{fig:stable}.  We should stress that for special values of $l$ there may exist additional instabilities due to resonances with the oscillations of the scale factor; such effects will be analyzed in \S \ref{sec:stability}.

There are some effects, however, which are not captured by (\ref{eq:massless}). In particular, we need to take into account that the fluctuations of the matter sources may have speed of sound $c_s \neq 1$, and that they also source gravitational fields. An important aspect here is that if the equation of state for the perfect fluid were to remain valid at short distances, the speed of sound $c_s^2=w<0$, and there would be instabilities at arbitrarily large momenta. Fortunately, as discussed in~\cite{usfirst}, this can be avoided by having solids that behave, on cosmological scales, as perfect fluids with negative $w$, but which nevertheless have $c_s^2>0$ and are stable. For the models discussed here, we need stable networks of domain walls and cosmic strings. See~\cite{Bucher:1998mh} for more details.

The effects from gravitational backreaction also turn out to be important. Incorporating the scalar metric perturbations and assuming a speed of sound $c^2_s > 0$ one obtains the following equation for the adiabatic mode, for the $l^{th}$ spherical harmonic (see e.g.\ \cite{usfirst}):
\beq
\psi'' + m^2_l \psi = 0 \, , \, m^2_l = \frac{(l+3)(l-1)c^2_s - 1}{K_{eff}}-\frac{7+9c^4_s}{4}\left(\frac{a'}{a}\right)^2 - \frac{3c^2_s - 1}{2}\frac{a''}{a}\,.
\eeq
To avoid introducing instabilities for the low-lying harmonics when $c^2_s(l+3)(l-1) <1$ we wish to keep $c_s$ an $O(1)$ fraction of the speed of light.\footnote{We also need to replace (\ref{stabilitylargebounce}) by the stronger condition $K_{eff} \ll c^2_s \gamma$.}  The $l=1$ spherical harmonic of the scalar metric perturbation is unstable due to a cancellation between terms in the equation of motion~\cite{usfirst} once we include backreaction. When there is only a single adiabatic perturbation sourced by one kind of matter this mode is pure gauge and can be redefined away; however, when multiple sources or fields are present -- as in our case -- this is no longer possible.  We can nevertheless project out this mode (and a few more of the low-lying spherical harmonics, if needs be) by orbifolding along the Hopf fiber of the $S^3$.  This orbifold will also remove Bianchi IX or Mixmaster perturbations, where the scale factor changes in a manner that is homogeneous but not isotropic.

\section{Stability and analytic considerations}\label{sec:stability}

Cyclic cosmologies have a potentially new type of instability that is not present in expanding universes. Due to the oscillatory behavior of the scale factor, it is possible to have resonances from modes whose momenta are close enough to integer multiples of the frequency of oscillation of the universe. 
In a flat infinite model there is a continuum of momentum modes and this always leads to instabilities. However, momenta on $S^3$ are quantized, and so it is possible that none of the modes lie within an instability band. This is the question that we now address.
Specifically, we will show that there exists a set of parameters $K_{eff}, \gamma$ for which there are no parametric resonances, and we discuss how the stability argument may be generalized beyond linear order using KAM theory.

In this section we focus, for simplicity, on the simple harmonic universe solution based on domain walls and cosmic strings; however, our conclusions are expected to be valid for the more general family of cosmologies discussed in \S \ref{subsec:general}.

\subsection{Hill's equation and parametric resonance}

Let us consider the stability of the solutions to the scalar field equation of motion (\ref{eq:psieom}), 
\beq
\psi'' + f(\hat{\eta})\psi = 0,
\eeq
where the forcing term for the $l^{th}$ spherical harmonic is
\be\label{eq:forcing1}
f(\hat{\eta})= \frac{l(l+2)}{K_{eff}} + \frac{2\sqrt{1-\gamma}\cos\hat{\eta}+(1-\gamma)(-3 + \cos 2\hat{\eta})}{2(1-\sqrt{1-\gamma}\cos\,\hat{\eta})^2}\,.
\ee
Recall that the rescaled conformal time is given by $\hat \eta= \sqrt{K_{eff}} \eta$, in terms of which the oscillating cosmology has period $2\pi$.
The generalization to the full equation for adiabatic perturbations including gravitational backreaction will be discussed later.

This class of equations, with a general periodic forcing term, are known as Hill's equations in mathematics, and are equivalent to a Schr\"odinger/Bloch wave problem in physics. A special and well-known case is the Mathieu equation, corresponding to $f(\hat{\eta}) = \delta + \epsilon \cos{\hat{\eta}}$. Stability of the solutions to Hill's equation has been extensively and rigorously studied; see for instance \cite{diffbook, hillandfourier, asymptoticbook}.  The solutions are known to be of the form
\beq
\psi(\hat \eta)=\psi_1(\hat{\eta})e^{\mu_1 \hat{\eta}} + \psi_2(\hat{\eta})e^{\mu_2 \hat{\eta}},
\eeq
where $\psi_1, \psi_2$ have period $2\pi$ and $\mu_1 \mu_2 = 1$.  More specifically, $\mu_1, \mu_2$ are either real reciprocals, in which case the solution is unstable without fine-tuning the initial conditions, or complex conjugates of unit magnitude, in which case the solution is stable for all time.  

Intuition from quantum mechanics suggests that the solutions should be stable in the limit where the forcing term changes adiabatically:
\beq
f'(\hat{\eta})/f(\hat{\eta})^{3/2} \ll 1,
\eeq
which holds in the static limit $(1-\gamma) \ll 1$ and in the large bounce limit as long as $K_{eff} \ll \gamma.$  More generally, when $K_{eff}/\gamma$ is large, the harmonics with $l \gg (K_{eff}/\gamma)^{1/2}$ are in the adiabatic regime.  This intuition is nearly correct; however, there will be `instability bands' in parameter space when the system has a parametric resonance between the free and driven oscillations.  To find a perturbatively stable bouncing universe, these instability bands must be avoided for all harmonics under consideration.

For a forcing term of the form
\beq\label{eq:generalpsi}
\psi'' + \omega(\omega - \frac{\epsilon}{\pi} p(\hat{\eta}))\psi = 0,
\eeq
where $ \int_0^{2\pi} p(x) dx = 1$, $\epsilon \ll 1$, as $\omega \to \infty$ the solution will pass through an infinite number of instability bands where the system approaches a parametric resonance point and becomes unstable.  Many asymptotic estimates exist for the width $\Delta \omega_n$ of the $n^{th}$ instability band -- in \cite{hillandfourier} it was shown that the width of the band is closely related to the $n^{th}$ Fourier coefficient of the forcing term $p(\hat{\eta})$.  Specifically, the solution will be unstable within a range
\beq
\label{intervals}
\omega_n = \frac{n}{2} + \frac{\epsilon}{2 \pi}(1 \pm |c_n|) + O(\epsilon^2)
\eeq
as long as the $n^{th}$ Fourier coefficient 
$$
c_n = \frac{1}{2\pi}\int^{2\pi}_0 p(\hat{\eta}) e^{-i n \hat{\eta}}d\hat{\eta}
$$ 
is nonzero.  For the forcing term in the simple harmonic universe, $c_n \to 0$ exponentially in $n$ as $n \to \infty$, as we will check below, and at first glance we might worry that the linear estimate for the size of the gap will eventually become smaller than the quadratic remainder term in \ref{intervals}.  However, it is easy to confirm using the proof in \cite{hillandfourier} that the $O(\epsilon^2)$ correction to the width of the instability interval remains subleading at large $n$.  More generally, when the forcing term is infinitely differentiable $\Delta \omega_n\,\to\,\frac{\epsilon}{\pi}|c_n|$ faster than any inverse power of $n$~\cite{asymptoticbook}.  The position of the instability band will also receive corrections order by order in powers of $\epsilon$ -- these, however, have a power-law falloff in $1/n$.

We will therefore use the estimate $\Delta \omega_n \approx \frac{\epsilon}{\pi}|c_n|$ to show that there exists a set of values for the parameters $K_{eff}, \gamma$ such that we can avoid the instability bands for all values of $n$ and $l$.   For the simple case of the Mathieu equation it is possible to give a constructive proof of stability, showing explicitly how to choose the values of $K_{eff}$ that avoid all parametric resonances --see e.g.~\cite{Auzzi:2012ca}. However, our forcing term is not just a cosine function and this complicates an explicit calculation of the stable values of $K_{eff}$.\footnote{One source of difficulties, absent in the Matthieu equation, is that the positions of the instability bands (\ref{intervals}) receive corrections in $1/n$ that cannot be neglected when estimating the closeness of $\omega$ to parametric resonances.} We will instead follow a different approach:
starting with a continuous set of $K_{eff}$ we discard the values that hit an instability band for the first spherical harmonic, then the ones that hit an instability interval for the second band, and so on, until we are left with the values that never hit an instability for any harmonic. We will show below that this process converges, and that the fraction of values of $K_{eff}$ that lead to instabilities is bounded and exponentially small both in the large bounce and quivering limits.

Before proceeding to our analysis of the large bounce and static limits, let us collect a few useful expressions. 
For the system at hand, the free oscillation is given by
\beq
\omega^2 = \frac{l(l+2)}{K_{eff}}\,,
\eeq
the parameter $\epsilon$ is equal to
\beq
\epsilon = \sqrt{\frac{K_{eff}}{l(l+2)}}\left(\frac{2\pi^2}{\sqrt{\gamma}} - 2\pi^2 \right)\,,
\eeq
and the normalized forcing term is given by
\beq
p(\hat \eta)=-\frac{1}{2\pi} \frac{\sqrt \gamma}{1- \sqrt \gamma}\frac{2\sqrt{1-\gamma}\cos\hat{\eta}+(1-\gamma)(-3 + \cos 2\hat{\eta})}{2(1-\sqrt{1-\gamma}\cos\,\hat{\eta})^2}\,.
\eeq
The $n^{th}$ Fourier coefficient is given by
\beq
c_{n} = -\frac{3(1 - \gamma)}{2}a_{n} + \frac{\sqrt{1-\gamma}}{2}(a_{n-1} + a_{n+1}) + \frac{1 - \gamma}{4}(a_{n-2} + a_{n+2})\, ,
\eeq
where
\beq
a_{n} = \frac{1}{\sqrt{\gamma} - 1}\left(\frac{\sqrt{1 - \gamma}}{1 + \sqrt{\gamma}}\right)^{|n|}\frac{(1 + |n|\sqrt{\gamma})}{\gamma}\,.
\eeq
For $|n|\ge 2$ this expression simplifies to
\beq\label{eq:cnapprox}
c_{n} = -\frac{2 |n| \sqrt{\gamma } -1 }{1-\sqrt{\gamma }}\left(\frac{\sqrt{1-\gamma }}{1+\sqrt{\gamma }}\right)^{|n|}\,.
\eeq

\subsection{Large bounce limit}

In the large bounce limit $\gamma \ll 1,$ we consider the regime $K_{eff} \ll \gamma$.  Since the instability intervals we have to avoid obey $n \gtrsim 1/\sqrt{K_{eff}}$, $n\sqrt{\gamma}$ is large.  Using (\ref{eq:cnapprox}), the leading term in the Fourier coefficient in this limit can be checked to be
\beq
c_n \rightarrow -2n \sqrt{\gamma}\,e^{-n\sqrt{\gamma}} + \,...
\eeq
The size of the $n^{th}$ instability interval is then
\beq
\Delta \omega_n \approx \frac{\epsilon}{\pi}|c_n| \sim \sqrt{\frac{K_{eff}}{l(l+2)}}\,n\,e^{-n\sqrt{\gamma}}\,.
\eeq
We see then that the lengths of the instability bands are exponentially small in the regime of interest $K_{eff} \ll \gamma$. We will next use this to argue that most values of $K_{eff}$ in such a range avoid all the resonant instabilities.

Consider $K_{eff}$ to be a continuous parameter varying within a narrow interval $I = \left[K^0_{eff}, K^0_{eff} + \delta K_{eff}\right]$ around a small value  $K^0_{eff}$.  By narrow we mean that $\delta K_{eff}/K^0_{eff} \ll 1$.  For a particular spherical harmonic $l = l_{*}$, as $K_{eff}$ varies over $I$, the function $\omega_{l_{*}}(K_{eff}) = \sqrt{\frac{l_{*}(l_{*}+2)}{K_{eff}}}$ will vary over a corresponding range $\left[ \omega_0, \omega_0 + \delta \omega \right]$ that includes potentially many instability bands.  The region $\delta_{l_{*}}K_{eff}$ of $I$ such that $\omega$ lies in an instability interval can be estimated using
\beq
\frac{\delta_{l_{*}}K_{eff}}{\delta K_{eff}} \sim \Delta \omega_n \,,
\eeq
which implies that the fraction of $I$ such that $\omega_{l_{*}}$ lies in an instability band is approximately
\beq
\frac{\delta_{l_{*}}K_{eff}}{\delta K_{eff}} \sim \Delta \omega_n \sim e^{-\sqrt{l_{*}(l_{*}+2)}\sqrt{\gamma/K^{0}_{eff}}}\,.
\eeq

The total fraction of the $K_{eff}$ that hit an instability band for some value of $l$ as $l$ ranges from $1$ to $\infty$ is therefore bounded by the expression
\beq
\sum_{l=1}^{\infty}\,e^{-\sqrt{l(l+2)}\sqrt{\gamma/K^{0}_{eff}}}
\eeq 
which is exponentially small as long as $K^{0}_{eff} \ll \gamma$.  Therefore in the parametric region $K_{eff} \ll \gamma$, most values of $K_{eff}$ will avoid instability bands for all $n$ and $l$, and the proof is done.

Note that while the fraction of $K_{eff}$ that avoid all parametric resonances is nearly one in a measure theoretic sense, the set of $K_{eff}$ that will hit an instability band eventually for some large but finite $l$ is dense, meaning that any neighborhood around a `good' choice will contain infinitely many bad choices.  Finding an explicit working set of values and assessing how `naturally' they are tuned depends on the context of the specific model used, while our aim here has been to show that there is a set of parameters that are free of resonant instabilities.  In practice, for the purposes of perturbative stability it suffices only to show stability for a finite (though potentially very large) set of modes, since for $l$ above a certain cutoff value $l_c$ the modes have Planck scale masses, in which case the classical instability will be replaced by nonperturbative particle or black hole production.  Under these relaxed criteria it is possible to find (exponentially small) open neighborhoods of the parameter $K_{eff}$ that avoid the finite collection of instabilities for $l < l_c$.

\subsection{Static limit}

As we discussed before, the static, or `quivering' limit $(1-\gamma) \ll 1$ is also conceptually interesting, especially for the possibility of constructing an eternal universe. Linearized fluctuations were found to be numerically stable in~\cite{usfirst} already with $K_{eff}=1$. We now show that it is also possible to avoid parametric resonances for most choices of parameters.

The Fourier coefficients of the normalized forcing term $p(\hat{\eta})$ in the static limit are given by
\be
c_n = -\frac{2}{1-\gamma}\,\left(\frac{\sqrt{1-\gamma}}{2}\right)^n (2n-1) + ...
\ee
for $n \geq 1$, and $c_0 = 1$.  The size of the $n^{th}$ instability band can therefore be estimated as
\beq
\Delta \omega_n \sim \frac{\epsilon}{\pi}|c_n| \sim \left(\frac{\sqrt{1-\gamma}}{2}\right)^n |(2n-1)|\sqrt{\frac{K_{eff}}{l(l+2)}}
\eeq
for $n \geq 1$, and $\Delta \omega_{n=0}\sim (1-\gamma)$.

Allowing $K_{eff}$ to vary in a narrow band $I = \left[K^0_{eff}, K^0_{eff} + \delta K_{eff}\right]$ around some small value $K^0_{eff}$, for a given mode $l = l_{*}$ the set $\delta_{l_{*}} K_{eff}$ of values for $K_{eff}$ such that $\omega_{l_{*}}^2(K_{eff}) = \frac{l_{*}(l_{*}+2)}{K_{eff}}$ lies within some nearby instability band can be estimated as
\beq
\frac{\delta_{l_{*}} K_{eff}}{\delta K_{eff}} \approx \left(\frac{\sqrt{1-\gamma}}{2}\right)^{\sqrt{l_{*}(l_{*}+2)/K^0_{eff}}}\,\left| 2\sqrt{\frac{l_{*}(l_{*}+2)}{K^0_{eff}}}-1\right|\,\sqrt{\frac{K^0_{eff}}{l_*(l_*+2)}}
\eeq
and so the total fraction of the set of $K_{eff}$ that hits an instability band for some value of $l$ as $l$ ranges from $1$ to $\infty$ is controlled by
\beq
\sum_{l = 1}^{\infty}\left(\frac{\sqrt{1-\gamma}}{2}\right)^{\sqrt{l(l+2)/K^0_{eff}}}\, \left|2-\sqrt{\frac{K^0_{eff}}{l(l+2)}}\,\right|\,
\eeq
which is exponentially small as long as $(1-\gamma) \ll 1$.  Once again, measure theoretically speaking, given a value of $\gamma$ in the quivering limit, most values of $K_{eff}$ will do the trick, and this time we do not need to worry about having $K_{eff} \ll 1$.  Actually, we do not even need the parameter $K_{eff}$ in this limit:\ treating $(1-\gamma)$ as the free parameter to be varied in a narrow range around a small value, the argument above carries through almost unmodified to show that for $(1-\gamma) \ll 1$ most values of $(1-\gamma)$ will avoid the entire infinity of resonances.


\subsection{Incorporating backreaction}

As we discussed in \S \ref{subsec:stability}, density perturbations from the domain walls and/or cosmic strings are not precisely described by the equation of motion for a free massless scalar. The main differences are that these fluctuations may have $c_s \neq 1$, and also metric perturbations become important. Let us now briefly comment on the stability for this case.

Incorporating gravitational backreaction on the scalar metric perturbations and assuming a speed of sound $c^2_s > 0$ for the adiabatic mode one obtains a Hill equation with forcing term
\beq
f(\hat{\eta}) = \frac{(l+3)(l-1)c^2_s - 1}{K_{eff}}-\frac{7+9c^4_s}{4}\left(\frac{a'}{a}\right)^2 - \frac{3c^2_s - 1}{2}\frac{a''}{a}\,.
\eeq
The exact value of $c^2_s$ depends on the shear resistance and is model-dependent -- but as we discussed in the previous section, in order to avoid introducing unstable modes when $c_s l \lesssim 1$ we wish to keep it an $O(1)$ fraction of the speed of light.  Then the arguments from before can be easily generalized to prove the existence of choices for the parameters $K_{eff}$, $\gamma$ such that we have stability in the large bounce and quivering limits.  

We can also generalize the discussion to give a more realistic treatment of the network dynamics, including effects such as shear modes, as occur in the solid dark matter model of \cite{Bucher:1998mh}. It is also straightforward to generalize the discussion to include isocurvature perturbations when multiple components are present.  However, we do not expect these points to qualitatively alter our conclusions.

\subsection{Nonlinearities}\label{subsec:nonlinear}

We have shown that we can have oscillating cosmologies that are stable at the linear level. It is possible that this stability is violated by non-linear couplings between the oscillating scale factor and the perturbative modes discussed above.  Much like the linear case, instabilities in the non-linear regime occur when a combination of the mode frequencies coupled together by a non-linear interaction fall into an instability band of the scale factor. When this occurs, the oscillations of the scale factor will resonantly excite these modes, leading to an instability. 

It is beyond our means to prove the eternal stability of our solutions in the presence of such non-linear instabilities.  But, for the following reasons, we believe that our solutions might be stable for a long time and may in fact be truly eternal at the linear and at the nonlinear level.\footnote{Nonperturbative decays are discussed below in \S \ref{sec:nonpert}.} A general way to study non-linear instabilities is to use KAM theory -- see e.g.~\cite{introkam} for a review.  The applicability of KAM theory though is limited by the requirement that the frequency of the excitable modes depend non-trivially on their amplitude. For a free theory, this condition is obviously not satisfied.  But, in an interacting theory, there will in general be non-linear interactions between these excitable modes (for example, a quartic interaction for a probe scalar).  At leading order, these interactions will not couple to the oscillating scale factor and are thus not a source of new instabilities. Their presence however will lead to a non-trivial dependence between the frequency of the mode and its amplitude, permitting the application of KAM theory. 

The KAM conditions boil down to ensuring that linear combinations of the excitable modes do not come sufficiently close to a resonant multiple of the oscillating scale factor. When these conditions are satisfied, KAM theory guarantees the eternal stability of the solution, even in the non-linear regime. Further, as shown in~\cite{Nekhoroshev}, when the KAM conditions fail, the non-linear solution will remain close to the (stable) linear solution for an exponentially long time, with a lifetime that depends upon the size of the failure of the KAM conditions.  Unfortunately, we do not have the expertise to mathematically perform the KAM analysis on our cosmological solutions. But this result suggests that even if our cosmology was not eternally stable, it should still survive for an exponentially long time. 

A physical reason to expect the demise of these cosmologies due to non-linearities is continuous entropy production.  This is simply a re-statement of the above mathematical issue -- the oscillations of the scale factor may resonantly excite modes, leading to the generation of entropy. The produced entropy will back react on the metric, potentially leading to runaway behavior.  But, these processes will not occur if these modes are far from the bands of instability of the system, and are thus subject to the results of a KAM-type analysis. In most physical systems, thermodynamic intuition suggests that the system will have these kinds of instabilities. Indeed, we except a disturbance of the system to dissipate through such excitations.  This intuition though is based on the fact that a typical thermodynamic system has a large number of modes that are nearly degenerate with the disturbance, making it easy for a linear combination of these frequencies to lie within the band of instability of the disturbance. 

Our oscillating cosmologies are interestingly different from such thermodynamic systems.  Here, we are interested in excitations caused by the oscillating scale factor which is the lowest frequency in the system. The potentially problematic excitable frequencies all correspond to higher harmonics on the sphere, making it easier to evade the degeneracies encountered in thermodynamic systems.  In fact, for sufficiently high frequencies, we expect notions of decoupling to become important -- a low frequency oscillation (such as that of the scale factor) cannot efficiently excite high frequency oscillations. This implies that the most problematic excitations will be low lying spherical harmonics that are roughly degenerate with the scale factor.  While it is unclear if these modes lead to instabilities, we note that they could be projected out by considering more complex geometries such as orbifolds of $S^3$ by a freely acting group.  Orbifolding will not change the equations of motion, but will project out modes from the spectrum (such a possibility was also considered in \cite{usfirst} and in \S \ref{subsec:stability} to project out linear instabilities). In particular, it will preserve the frequency of the oscillating scale factor but can eliminate low lying modes that are roughly degenerate with it. The resulting system is expected to be more stable since the potentially problematic modes are all at high frequencies compared to that of the scale factor.  Another possibility is to introduce additional sources of energy density such as a frustrated network of strings. As we have seen in \S \ref{sec:multifluid}, these sources can parametrically make the frequencies of the fluctuations higher than the frequency of the scale factor. This should also enhance the non-linear stability of the system. It is an interesting open question whether such strategies could produce a cosmology that is eternal to all orders in perturbation theory.

\section{Comments on nonperturbative decays}\label{sec:nonpert}

A simple harmonic universe that is stable at the linear and nonlinear level may still
be vulnerable to nonperturbative decays on exponentially long timescales~\cite{usfirst}. An interesting example of such a process was found by Mithani and Vilenkin~\cite{Mithani:2011en}, who used the effective potential of the scale factor to construct an instanton that mediates tunneling to zero size.  Here we wish to add a few comments on this decay channel and show how it can be avoided using well-controlled corrections that become important at short scales. 

The main point is that the endpoint of the instability is singular and both its amplitude and its existence may be sensitive to the physics operating near $a \sim l_{Pl}$. While~\cite{Mithani:2011en} assumed that the effective GR description is valid all the way up to the Planck scale, it is very plausible that other effects (higher derivative terms, quantum corrections, etc.), which become important before reaching $M_{Pl}$, will modify the Mithani-Vilenkin instanton. We will provide a simple example of this, by showing that the Casimir energy from a collection of scalar fields associated to the compact $S^3$ of the SHU will lift that instanton, in a controlled regime where our semiclassical approximation is valid. We will also comment briefly on other possible decay modes.

Let us begin by reviewing the solution of~\cite{Mithani:2011en} and then show that it can be avoided very simply using Casimir energy from scalar fields. In the minisuperspace approximation, the Wheeler-de Witt wavefunction for the SHU obeys (ignoring ordering ambiguities)
\beq
\left(-\frac{d^2}{da^2} + V(a)\right)\psi(a) = 0
\eeq
where
\beq\label{eq:Va2}
V(a) = \left(\frac{3\pi}{2 G_N}\right)^2 a^2 \left(K_{eff} - \frac{8\pi G_N}{3}(c_{dw}\, a + \Lambda a^2)\right)\,.
\eeq
This follows from (\ref{eq:GRH}) and (\ref{eq:Va}).

Setting $K_{eff} = 1$ and tuning for simplicity parameters $\{ c_{dw}, \Lambda \}$ so that the Lorentzian solution is in the static limit $a_{stat} = \frac{1}{\omega}$, the Euclidean Friedmann equations admit the solution 
\beq\label{eq:instanton1}
a(\tau) = \frac{1}{\omega}(1 - e^{-\omega \tau})
\eeq
\noindent and the Euclidean action is given by
\beq
|S_{inst}| = \int^{\infty}_0 d\tau a[\dot{a}^2 + (\omega a - 1)^2] = \frac{3}{32 G_N^2 |\Lambda|}.
\eeq
This assumes that (\ref{eq:Va2}) is valid even for very small values of $a$.
Mithani and Vilenkin interpret this as a tunneling amplitude to zero scale factor, and check that the amplitude agrees with the WKB amplitude one calculates from $V(a)$:
\beq
\int_0^{a_{stat}} \sqrt{V(a)} da = \frac{3}{32 G_N^2 |\Lambda|}
\eeq

The solution is singular at the point of vanishing scale factor, and will be sensitive to the physics resolving the singularity as well as to other sources and corrections which may become important before reaching a Planckian size.  In particular a contribution to the energy such that $\delta V(a) \propto a^{n \leq 0}$ would lift the minimum of $V(a)$ at zero scale factor and prevent the Mithani-Vilenkin process from taking place. It is interesting that for the SHU there is a very natural contribution of this type, namely the Casimir energy associated to the compact spatial slices. More specifically, the Casimir contribution from $N$ massive scalar fields has the right parametrics (and the correct sign, as checked e.g. in~\cite{Herdeiro:2005zj}) to accomplish this:
\beq
\delta V(a)\sim \left(\frac{3\pi}{2 G_N}\right)^2 a^2 \left(\frac{8\pi G_N}{3}\frac{N}{a^2}\right)
\eeq
We have included multiple fields so that when the Casimir energy begins to dominate $V(a)$ at around $a \sim \sqrt{8 \pi G_N N}$, we can still be well within the semiclassical regime as long as $N \gg 1$. This contribution to the effective potential lifts the tunneling instability (\ref{eq:instanton1}). Furthermore, the results of \S \ref{sec:stability} imply that the introduction of these additional scalars does not lead to new linearized instabilities.

In the presence of Casimir energy, there still exists a Euclidean solution interpolating between the static universe and $a = 0$.\footnote{We thank A.~Guth and A.~Vilenkin for pointing this out to us.} This solution has finite action but, unlike (\ref{eq:instanton1}), now the instanton has a singular geometry with $a(\tau) \propto \tau^{1/2}$ as the Euclidean time $\tau \to 0$. It is therefore not clear that this solution should be allowed.\footnote{As a simpler example, in gauge theories only very special singularities are allowed, such as certain singular monopoles from 't Hooft lines.} The question of which singularities are physical in GR is an important open problem that requires a theory of quantum gravity. Without such an understanding here, the standard approach of only path-integrating over smooth configurations implies that this instanton does not contribute. This illustrates again our main point that discussing the behavior near $a \to 0$ requires a UV complete theory of gravity, where various new effects are expected to become important at or before Planckian scales. As another example, higher order curvature corrections may become important as well, and it would be interesting to understand which way these can push.

A smiliar set of questions could be asked of the milder singularity appearing in the solution (\ref{eq:instanton1});\ of course, in this case it is also possible to imagine scenarios where this can be resolved and the decay process can be present. For instance,~\cite{Mithani:2011en} proposed that the singularity may be a coordinate singularity from the projection of a higher-dimensional model~\cite{Garriga:1998ri}, in which case the UV completion comes from the physics of the extra dimensions and need not affect the amplitude of the four-dimensional instanton.  The solution can also be regularized by adding a small amount of dust or radiation, such that the endpoint of the tunneling process occurs at small but nonzero scale factor; the instanton is no longer singular, but the amplitude will be nearly unchanged~\cite{Dabrowski:1995jt, Mithani:2011en}.

On the other hand, there may also be other nonperturbative processes that dominate the decay rate. For instance, these may involve black hole nucleation or collapse of the metastable string and/or wall network. While the rate of these processes is model-dependent, we would expect them to progress more rapidly than the gravitational decay modes. It would be interesting to explore the rates and tunabilities of various decay modes in the context of an explicit model, either in field theory or in string theory.  

In any case, we believe that the modification of the SHU that we just described, together with the seemingly special thermodynamic properties of the static limit, make it a promising candidate for exploring the theoretical limits on the (meta)stability of an eternal universe, an issue to which we hope to return in the future.

\section{Conclusions and future directions}\label{sec:concl}

In this work we have extended the model and results of~\cite{usfirst} to allow for solutions with large hierarchies between the minimum and maximum sizes of the universe, and studied their stability properties. Stable oscillating universes with a large hierarchy were obtained by combining a network of frustrated cosmic strings, matter with $-1 < w < 1/3$, a negative cosmological constant, and positive curvature. We have found that these cosmologies can be made fully stable at the linearized level, and we ruled out the possibility of parametric resonances in a large range of parameter space.  A point worth stressing once again is that while for analytic simplicity we have focused on the simple harmonic universe (SHU), where a frustrated network of domain walls contributes $w = -2/3$, we expect our results can be extended to this more general class of models.

The simple harmonic universe could suffer from non-perturbative instabilities, which would lead to a metastable and exponentially long-lived, though not eternal, universe.  While there are processes that lead to such instabilities \cite{Mithani:2011en}, we have shown that the simplest of these can be cured through suitable modifications of short distance physics that are naturally present in our model.  We have not ruled out the possibility of non-perturbative decay more generally, but it is not yet clear to what extent this question is model-dependent, and whether the limitations are universal or specific.  A broader class of instabilities may arise from continuous entropy production.  While we were unable to show that these instabilities do not exist in our model, we also could not find concrete arguments that establish their existence.  It would be interesting to further study these classes of instabilities.   

It would also be instructive to develop further a microscopic realization of the simple harmonic universe, in the context of string theory or even at the level of field theory.  Furthermore, while have not pursued the inclusion of inflation and the hot Big Bang phase into the simple harmonic universe, it would definitely be interesting to do so. Decays of the SHU into inflating universes may also lead to new possibilities in the string landscape, an alternative that we plan to analyze in the future. It would also be worth pursuing more general questions regarding the structure of GR, in particular whether there are quantum counterparts of the singularity theorems, or more general notions about the arrow of time and the definition of entropy in such a setup.  

Finally, it is of interest to understand whether an eternal universe may be possible on general grounds -- see \cite{Wilczek} for an interesting discussion of spontaneous breaking of time translation at the classical and quantum level.  Recent work arguing for the existence of a past singularity and a `beginning of time' \cite{Mithani:2011en, Susskind:2012yv, Mithani:2014jva} in the context of the universe has focused on the generation and role of entropy or on the presence of specific decay modes.  It is not clear, however, that these arguments apply to our class of solutions, particularly in the static limit and with the addition of Casimir energy described in \S \ref{sec:nonpert}.  We hope that our study of these models will stimulate further discussion of these issues.

\section*{Acknowledgements}

We especially thank Shamit Kachru for early collaboration on these and related issues, and we further thank Adam Brown, Xi Dong, Sergei Dubovsky, Brian Greene, Alan Guth, Shamit Kachru, Ali Masoumi, Audrey Mithani, Alex Vilenkin, and Erick Weinberg for very helpful discussions.  PWG, BH, and GT thank the KITP for hospitality during the program ``Primordial Cosmology,'' during which a portion of this work was completed.  PWG~acknowledges the support of NSF grant PHY-1316706, the Hellman Faculty Scholars program, and the Terman fellowship.  BH~is supported in part by the DOE Office of Science under grant DE-FG02-92-ER40699.  SR~was supported by ERC grant BMSOXFORD no.\ 228169.  GT~is supported in part by the National Science Foundation under grant no.~PHY-0756174.


\bibliographystyle{JHEP}
\renewcommand{\refname}{Bibliography}
\addcontentsline{toc}{section}{Bibliography}
\providecommand{\href}[2]{#2}\begingroup\raggedright

\end{document}